\documentclass[orivec]{llncs}

\usepackage[T1]{fontenc}
\usepackage[pdftex]{graphicx}
\usepackage[font=footnotesize]{caption}
\usepackage{cmap}
\usepackage{bm}
\usepackage{amsmath,amssymb,mathabx}
\usepackage[dvipsnames]{xcolor}
\usepackage{fancyvrb}
\usepackage{inconsolata}
\usepackage{array}
\usepackage{subfig}
\usepackage{float}

\definecolor{lightgray}{gray}{0.7}

\let\textquotedbl="

\def\ExclamationPoint{\char59}
\catcode`;=\active

\DefineVerbatimEnvironment{VerbOutC}{Verbatim}{fontfamily=zi4}

\RecustomVerbatimCommand{\VerbatimInput}{VerbatimInput}%
{ 
 defineactive=\def;{\color{lightgray}\ExclamationPoint},
 framerule=0.25mm,
 numbers=left,
 frame=lines,  
 framesep=2em, 
 labelposition=topline,
}

\newcommand{\rosette}{\textsc{Rosette}}
\newcommand{\runb}{\textsc{Rosette}/\textsc{Unbound}}

\newcommand{\mochi}{\textsc{MoCHi}}
\newcommand{\spacer}{\textsc{Spacer3}}
\newcommand{\muz}{\textsc{Z3}}

\newcommand{\hcs}[1]{{\scriptsize\[\left\{{#1}\right.\]}}

\newcommand{\para}[1]{\smallskip\noindent\textbf{#1}}

\newcolumntype{C}{>{\centering\arraybackslash}p{1cm}}

\begin{document}

\title{Verifying Safety of Functional Programs with \runb}

\author{Dmitry Mordvinov\inst{1} \and Grigory Fedyukovich\inst{2}}

\institute{Saint-Petersburg State University, Department of Software Engineering, Russia, \\
\email{dmitry.mordvinov@se.math.spbu.ru}
\and University of Washington Paul G. Allen School of Computer Science \& Engineering, USA,
\email{grigory@cs.washington.edu}}

\maketitle

\pagestyle{plain}

\begin{abstract}

The goal of unbounded program verification is to discover an inductive invariant that safely over-approximates all possible program behaviors.
Functional languages featuring higher order and recursive functions become more popular due to the domain-specific needs of big data analytics, web, and security.
We present \runb, the first program verifier for Racket exploiting the automated constrained Horn solver on its backend.
One of the key features of \runb{} is the ability to \emph{synchronize} recursive computations over the same inputs allowing to verify programs that iterate over unbounded data streams multiple times.
\runb{} is successfully evaluated on a set of non-trivial recursive and higher order functional programs.

\end{abstract}
\section{Introduction}

Rapid growth in data sciences, web, and security opens the new dimensions for functional programming languages~\cite{DBLP:journals/dagstuhl-reports/GaboardiJJW16}.
Due to streaming processing over ``big data'', applications to program synthesis and to the design of domain specific languages (DSLs) impose extra safety requirements in the unbounded setting.
This makes the ability to discover inductive invariants crucial and necessitates the proper support by the verification tools.
We present \runb{}, a new formal verifier for Racket%
\footnote{\url{https://docs.racket-lang.org/}}%
, whose most distinguishing feature is the tight connection to the solver of constrained Horn clauses (CHC) offering an automated decision procedure aiming to synthesize safe inductive invariants.

\runb{} is available in the interactive mode of the DrRacket IDE%
\footnote{\url{https://docs.racket-lang.org/drracket/}}%
, allowing the users to verify the programs without detracting from  coding.
It uses a symbolic execution engine of the  \rosette~\cite{DBLP:conf/oopsla/TorlakB13,DBLP:conf/pldi/TorlakB14} tool (thus, sharing a part of the name with it), but it implements a conceptually different encoding strategy and relies on a conceptually different constraint solving paradigm.
Furthermore, it features a support of unbounded symbolic data type of lists, which would be impossible in \rosette{} due to its bounded nature, but is still compatible with the existing bounded reasoning.

Yet another key feature of \runb{} is the ability to verify multiple-pass list-manipulating programs through automatic deriving so called \emph{synchronous} iterators.
While treating each element of a list nondeterministically, our tool ensures that it is accessed by all list iterators exactly once and exactly at the same time. 
We empirically show that this program transformation allows to effectively verify non-trivial safety properties (e.g., the head of a sorted list of integers is always equal to its minimal element).
To the best of our knowledge, \runb{} is currently the only tool featuring such functionality.

The paper is structured as follows.
Sect.~\ref{sec:impl} describes the workflow of \textsc{Rosette}/ \textsc{Unbound} and lists the most important features of the tool, which allow it to verify a program in Sect.~\ref{sec:ex}.
Sect.~\ref{sec:eval} reports on the empirical evaluation of the tool, and finally Sect.~\ref{sec:related} outlines the closest tools and concludes the paper.

\section{Tool Overview}
\label{sec:impl}

From a distance, \runb{} implements the workflow exploited by the SMT-based unbounded model checkers~\cite{DBLP:conf/fm/HojjatKGIKR12,DBLP:conf/cav/KomuravelliGC14,seahorn,DBLP:conf/cav/KahsaiRSS16,DBLP:journals/corr/GarocheGK14,DBLP:journals/corr/GarocheKT16,DBLP:conf/pldi/KobayashiSU11}.
As an intermediate representation of the \emph{verification condition}, it uses a system of constraints in second order logic which encode the functional program in Racket including the assertion specifying a safety property.
While the functionality of solving the system of constraints is entirely due to the external solver, \runb{} contributes in an efficient way of constructing the solvable systems.

Like \rosette{}, the first steps of \runb{} are the symbolic execution of the program, symbolic merging of states in the points where branches of the control flow join~\cite{DBLP:conf/pldi/TorlakB14}, and transforming the user code by adding some system routines' calls for tracking its execution.
This results in a compact symbolic encodings of separated acyclic parts of the program (thus acting close enough to the Large Block Encoding~\cite{DBLP:conf/fmcad/BeyerCGKS09}) and ensures that no information about them is lost.
Then, \runb{} uses the individual encodings for generating concise systems of constrained Horn clauses (CHCs).
Unlike \rosette{}, our tool explicitly maintains the call graph and a set of uninterpreted relation symbols $R$, and for each function $f$ that has not been encoded yet, it creates a fresh symbol $f_r$ and inserts it to $R$.

For every (possibly merged) execution branch of $f$, called with $\vec{in}$, the tool creates the following implication,  the premise of which is  the conjunction of first-order path conditions $\phi$ and the nested function calls $f_i$ outputting the values $\vec{out}_i$ (i.e., return values and/or state mutations) produced in this branch for $\vec{in}_i$:
\begin{equation}
\label{eq:chc}
	f_{r}(\vec{in}, \vec{out})\text{,} \gets \phi , f_{r_1}(\vec{in}_1, \vec{out}_1) , \ldots, f_{r_n}(\vec{in}_n, \vec{out}_n) 
\end{equation}
%
The verification condition with assumptions $pre$ and requirements $post$ for a piece of program calling functions $f_1, \ldots, f_m$, is translated into:
\begin{equation}
\label{eq:ass}
	false \gets pre, f_{r_1}(\vec{in}_1, \vec{out}_1) , \ldots , f_{r_m}(\vec{in}_m, \vec{out}_m) , \neg post
\end{equation}

We refer to an implication of form either~\eqref{eq:chc} or~\eqref{eq:ass} as to constrained Horn clause.
A system of CHCs serves as a specification for synthesis of inductive invariants, and it is called solvable if there is an interpretation of symbols in $R$ making all implications~\eqref{eq:chc} or~\eqref{eq:ass} true. \runb{} uses off-the-shelf Horn solver to obtain solutions for the system of CHCs it produces. The default Horn solver used by \runb{} is \spacer{}~\cite{DBLP:conf/cav/KomuravelliGC14}, but it is also possible to switch to \muz{}~\cite{DBLP:conf/sat/HoderB12}.

In the rest of the section, we describe the implementation details and the outstanding features of \runb{}. 

\para{Multiple user-specified assertions.}
Each Racket function may contain numerous implicit and explicit assertions. 
The call of \texttt{verify/unbound} at some point of symbolic execution creates a system of CHCs by encoding all possible program executions up to that point as~\eqref{eq:chc}, and encodes so called \emph{main assertions} (specified with arguments of \texttt{verify/unbound}) as~\eqref{eq:ass}.
\runb{} gradually propagates assertions from the function bodies and conjoins them with the main assertions:
thus, to find a bug, our tool needs to find a violation of at least one of all assertions.

\para{Mutations and global variables.}
Global state accessed and mutated in the body of function $f$ is added as an extra argument to relation $f_r$. 
\textsc{Rosette}/ \textsc{Unbound} obtains a symbolic encoding of the mutations performed by arbitrary (possibly, mutually recursive) functions. 
One important feature is that unused global state is excluded from the encoding, keeping relations compact even for large programs.

\para{Higher order functions.}
Racket allows manipulating functions as arguments to another functions, the feature common to functional programming languages. 
To encode a higher order function to CHCs, \runb{} needs a particular assignment $A$ to its arguments and an assignment to the arguments of all other higher order function appearing among $A$.
This way, \runb{} performs a so called \emph{higher order inlining} whose goal is to instantiate all functional arguments with their particular meaning. 
Thus, each new application of a higher order function is treated as a unique new function, necessitating the creation of a fresh uninterpreted relation symbol in $R$ and constructing a set of separate CHCs.

\para{Unbounded symbolic lists.}
\runb{} introduces the unbounded symbolic data type of lists and supports built-in operations over lists (including the length, head, tail, iterators, mapping and appending functions whose outputs are the tailored symbolic constants).
However, \runb{} also supports a partially specified lists (e.g., it allows to cons a symbolic list and a concrete head).
The data type of symbolic lists is useful while verifying properties about list iterators (e.g, using a higher order function $fold$).
That is, while an element is accessed in each iteration over the symbolic list, it is treated nondeterminstically. 
It has its negative side effect while dealing with multiple traversals over the same list.
Indeed, nondeterminism relaxes the fact that both traversals are conducted over the same list. 

\para{Nondeterminism modulo synchronization.}
We compensate this weakness by merging the individual iterators over lists and producing a new CHC system to be solved instead of the original one.
In the lower level, it requires adding the extra synchronization constraints that equates the elements at the same position of the same list while being accessed by different iterators.
The similar reasoning is exploited for proving relational properties over numeric recursive programs (e.g., monotonicity of the factorial).
This transformation is proven sound in our previous work~\cite{MF_LPAR_2017}.


\section{Running Example}
\label{sec:ex}

\newcommand{\fld}{\textsc{sum}}
\newcommand{\fldm}{\textsc{sum$_{_{+1}}$}}
\newcommand{\flds}{\textsc{sums}}

\begin{figure}[t!]

\begin{Verbatim}
#lang rosette/unbound
; functions over integers:
(define/typed (inc/typed x) (~> integer? integer?) (+ x 1))
(define/typed (+/typed x y) (~> integer? integer? integer?) (+ x y))
; nondeterministically treated list:
(define-symbolic xs (listof integer?))
; assertion requiring multiple list traversals:
(verify/unbound (assert (= (+ (foldl +/typed 0 xs) (length xs))
                           (foldl +/typed 0 (map inc/typed xs)))))
\end{Verbatim}
\hcs{
  \begin{aligned}
    \fld (\ell_1, acc_1, res_1)  \gets & \ell_1 = 0, res_1 = acc_1 \cr
    \fldm (\ell_2, acc_2, res_2)  \gets & \ell_2 = 0, res_2 = acc_2 \cr
    \fld (\ell_1, acc_1, res_1)  \gets & \fld (\ell_1 -1 , acc_1, res'_1) , \ell_1 > 0, res_1 = res'_1 + hd_1 \cr
    \fldm (\ell_2, acc_2, res_2)  \gets & \fldm (\ell_2 -1 , acc_2, res'_2) , \ell_2 > 0, res_2 = res'_2 + hd_2 + 1 \cr
    \bot \gets & \fld(\ell, 0, res_1) , \fldm (\ell, 0, res_2) , \ell \ge 0 , res_1 + \ell  \neq res_2
  \end{aligned}
}

\noindent\rule[0.5ex]{\linewidth}{0.75pt}

\hcs{
  \begin{aligned}
    \flds (\ell_1, acc_1, res_1, \ell_2, acc_2, res_2) \gets & \ell_1= 0, res_1 = acc_1 , \ell_2= 0, res_2 = acc_2  \cr
    \flds (\ell_1, acc_1, res_1, \ell_2, acc_2, res_2) \gets & \flds (\ell_1 -1 , acc_1, res'_1, \ell_2 -1 , acc_2, res'_2) , \bm{hd_1 = hd_2} \cr
    		& \ell_1 > 0, \ell_2 > 0, res_1 = res'_1 + hd_1, res_2 = res'_2 + hd_2 + 1 \cr
    \bot \gets & \flds(\ell, 0, res_1, \ell, 0, res_2) , \ell \ge 0 , res_1 + \ell  \neq res_2
  \end{aligned}
}

\noindent\rule[0.5ex]{\linewidth}{0.75pt}

\caption{Higher order verification with \runb: a given program, the intermediate and ultimate encoding.}%
\label{ex:1}
\end{figure}

We illustrate the \runb{} workflow on an example.
Consider a higher order functional program in Fig.~\ref{ex:1} (upper).
It features two lists of integers: \texttt{xs} and \texttt{(map inc/typed xs)}.
The latter is created from \texttt{xs} by adding \texttt{1} to each element of \texttt{xs}.
Then, the program performs two traversals and sums all the elements of each list: \texttt{(foldl +/typed 0 $\ldots$)}.
We want to verify that the difference between the results of those summing iterations is exactly equal to \texttt{(length xs)}.

\runb{} creates a verification condition as a system of CHCs  over linear arithmetic.
There are two main encoding stages:
\begin{enumerate}
  \item Creating individual CHCs for all list iterators:
    	   \begin{itemize}
	       \item $\texttt{(foldl +/typed 0 xs)} \mapsto \fld(\ell_1, acc_1, res_1)$,
	       \item $\texttt{(foldl +/typed 0 (map inc/typed xs))} \mapsto \fldm(\ell_2, acc_2, res_2)$.
	   \end{itemize}
	   Each call to a higher order function is ``inlined'' into a set of CHCs.
	   Despite there are two calls of the same function \texttt{fold}, they are encoded separately as they get different functions as parameters.
	   Note that the other higher order functions (e.g., \texttt{map}), if they are used only as parameters to some other higher order functions, are not encoded as separate CHCs.
	   The system obtained as a result after this step is shown in Fig.~\ref{ex:1} (central).
  \item Synchronizing the CHCs using \texttt{xs}:
    	   \begin{itemize}
           	\item {\footnotesize$\fld(\ell_1, acc_1, res_1) \!\Join \!\fldm(\ell_2, acc_2, res_2) \!\mapsto\! \flds (\ell_1, acc_1, res_1,\ell_2, acc_2, res_2)$}.
           \end{itemize}
           Importantly, in this step, an equality over lists heads is implanted to the constructed system.
           The ultimate system (with the implanted equality in bold) is shown in Fig.~\ref{ex:1} (lower).
           The key insight behind the system is that all reasoning is reduced to a single computation over a single nondeterministic list: each iteration accessing the head is shared among the original iterators.
\end{enumerate}
Finally, \runb{} automatically passes the ultimate system of CHCs to the external solver and waits while the solving is delivered.

\section{Evaluation}
\label{sec:eval}

We evaluated our tool on a set of functional programs challenging for verification, i.e., 
(1) programs performing computations in non-linear arithmetics (e.g., monotonicity of powers, relational properties of \texttt{div}, \texttt{mod}, and \texttt{mult} operations, etc.) expressed as recursive functions using only linear arithmetics,
(2) programs with recursive and mutually recursive functions performing mutations of variables,
(3) programs with higher order functions, and
(4) programs asserting complex properties of a nondeterministic symbolic list, due to its multiple (explicit or implicit) traversals.

\para{Comparison with \mochi.} 
Since there is no unbounded verifier for Racket, we compared our tool with \mochi{}~\cite{DBLP:conf/pldi/KobayashiSU11}, a state-of-art verifier of OCaml programs.
Both tools also use constrained Horn solving in their verification backend. 
We translated the benchmark set of \mochi{} to Racket and symmetrically translated our benchmark set to OCaml. 
The common features of \runb{} and \mochi{} include the support of programs with recursion, linear arithmetics, lists.
However, contrary to the current revision of \runb{}, \mochi{} supports algebraic data types and exceptions handling.
Thus, making \textsc{Rosette}/ \textsc{Unbound} work for the remaining \mochi's benchmarks is left for our future work.

All benchmarks that we managed to translate for our tool were correctly solved. 
On the other hand, 6 of our benchmarks contain programs with mutations that not yet supported by \mochi{}. 
For 9 of 12 supported programs with bugs, \mochi{} managed to find a counterexample.
However, only for 1 of 12 safe programs, \mochi{} managed to prove its correctness (which is significantly harder than searching for a finite counterexample).
For 11 others,  either a timeout was reached or an internal error occurred.
Table~\ref{tbl:shortStats} shows the brief statistics of our evaluation.%
\footnote{The complete table can be found in Appending~\ref{append:data}.}

\begin{table}[t!]
\begin{center}
\begin{tabular}{ |l|l|l| } 
 \hline
               & \runb's benchmarks               & \mochi's benchmarks \\ \hline
 \runb{}       & 30/30, ~0.087/0.115 sec   & 40/54, ~0.075 sec      \\ \hline
 \mochi        & 10/30, ~0.475 sec            & 54/54, ~0.227/0.590 sec       \\ 
 \hline
\end{tabular}
\end{center}
\caption{The overall statistics. A ratio in the first position of each cell corresponds to the solved and the total numbers of benchmarks from the corresponding set. The first value in the second position stands for the average time of solving the opponent's benchmarks. The second value (if any) in the second position stands for the average time on all benchmarks.}
\label{tbl:shortStats}
\end{table}

\section{Closing Remarks}
\label{sec:related}

Verification of functional languages is typically based on type inference~\cite{DBLP:conf/popl/OngR11,DBLP:conf/icfp/KakiJ14,DBLP:conf/icfp/VazouSJVJ14,DBLP:conf/popl/SwamyHKRDFBFSKZ16,DBLP:conf/flops/0001KSP16}.
Our tool implements an orthogonal, SMT-based approach which is native also for the Dafny~\cite{DBLP:conf/lpar/Leino10}, Leon~\cite{DBLP:conf/sas/SuterKK11}, and Zeno~\cite{DBLP:conf/tacas/SonnexDE12} induction provers.
None of those approaches has an completely automated CHC solver on its backend.
In contrast, CHC solvers are exploited by the model checkers for imperative languages~\cite{DBLP:conf/fm/HojjatKGIKR12,DBLP:conf/cav/KomuravelliGC14,seahorn,DBLP:conf/cav/KahsaiRSS16}, dataflow languages~\cite{DBLP:journals/corr/GarocheGK14,DBLP:journals/corr/GarocheKT16}, and functional programming languages~\cite{DBLP:conf/pldi/KobayashiSU11}.
Our work makes another functional programming language, rapidly becoming more popular, communicate with an external CHC solver.

One of the main contributions of our \runb{} is the support of new unbounded data type of symbolic lists.
Furthermore, the ability to implicitly merge recursive functions over the same input data make \textsc{Rosette}/ \textsc{Unbound} the first program verifier which is able to successfully deal with programs that iterate over unbounded data streams multiple times.
To confirm this claim, \runb{} was successfully evaluated on a set of non-trivial recursive and higher order functional programs.


Applying \runb{} to functional program synthesis~\cite{DBLP:conf/cav/AlbarghouthiGK13,DBLP:conf/cav/KneussKK15,DBLP:conf/pldi/PolikarpovaKS16,grasspWorkshop}
is the next step of our future work.
Indeed, the support of multiple-pass list manipulating programs also enables writing ``universally quantified'' program specifications.
It would be also interesting to see how the discovered inductive invariants could be encoded as the first class entities and be a part of the executable Racket code.

\bibliographystyle{abbrv}
\bibliography{references}

\newpage
\appendix
\section{Raw Experimental Data}
\label{append:data}

\begin{figure}[h]%
\vspace{-3em}
\centering%
\resizebox{0.91\linewidth}{!}{%
\scriptsize%
\subfloat[on \mochi{} benchmarks~\cite{DBLP:conf/pldi/KobayashiSU11}]{%
\begin{tabular}{ |p{0.3\linewidth}|c|C|C| } %
 \hline %
 \emph{benchmark name}	 	& \mochi 	& {\tiny\textsc{Rosette}/ \newline \textsc{Unbound}} \\ \hline %
ack			 	& 0.096			& 0.063 	\\ \hline  %
a-cppr 			& 2.298			& NI 	\\ \hline  %
a-init			& 3.441			& NI 	\\ \hline  %
a-max-e			& 0.684			& 0.063 	\\ \hline %
a-max			& 0.687			& 0.062 	\\ \hline %
copy\_intro		& 0.159			& 0.177 	\\ \hline %
e-fact			& 0.096			& 0.058 	\\ \hline %
e-simple		& 0.060			& 0.017 	\\ \hline %
fact-not-pos-e	& 0.067			& NI 	\\ \hline %
fact-not-pos 	& 0.110			& NI 	\\ \hline %
fold\_div-e		& 0.084			& 0.118 	\\ \hline %
fold\_div		& 0.152			& 0.136 	\\ \hline %
fold\_fun\_list	& 0.114			& NI 	\\ \hline %
fold\_left		& 0.224			& 0.085 	\\ \hline %
fold\_right		& 0.160			& 0.085 	\\ \hline %
forall\_eq\_pair	& 0.417			& NI 	\\ \hline %
forall\_leq		& 0.398			& 0.091 	\\ \hline %
harmonic-e		& 0.078			& 0.146 	\\ \hline %
harmonic		& 0.144			& 0.141 	\\ \hline %
hors			& 0.072			& 0.092 	\\ \hline %
hrec 			& 0.102			& NI 	\\ \hline %
intro1			& 0.059			& 0.031 	\\ \hline %
intro2			& 0.060			& 0.031 	\\ \hline %
intro3			& 0.058			& 0.021 	\\ \hline %
isnil			& 0.121			& 0.054 	\\ \hline %
iter			& 0.122			& 0.091 	\\ \hline %
length			& 0.141			& 0.059 	\\ \hline %
l-zipmap		& 0.104			& 0.118 	\\ \hline %
l-zipunzip 		& 0.119			& NI 	\\ \hline %
a-max			& 0.068			& 0.079 	\\ \hline %
mc91-e			& 0.078			& 0.049 	\\ \hline %
mc91			& 0.441			& 0.056 	\\ \hline %
map\_filter		& 10.951		& NI 	\\ \hline %
map\_filter-e	& 0.934			& NI 	\\ \hline %
mem			 	& 0.191			& 0.090 	\\ \hline %
mult-e			& 0.096			& 0.048 	\\ \hline %
mult			& 0.147			& 0.055 	\\ \hline %
neg			 	& 0.066			& 0.022 	\\ \hline %
nth0			& 0.175			& 0.095 	\\ \hline %
nth			 	& 0.534			& 0.115 	\\ \hline %
repeat-e		& 0.065			& 0.050 	\\ \hline %
reverse			& 0.281			& 0.102 	\\ \hline %
risers			& 1.435			& NI 	\\ \hline %
r-file			& 1.476			& 0.135 	\\ \hline %
r-lock-e		& 0.070			& 0.082 	\\ \hline %
r-lock			& 0.062			& 0.089 	\\ \hline %
search-e		& 0.385			& 0.028 	\\ \hline %
search			& 0.659			& 0.032 	\\ \hline %
sum-e			& 0.064			& 0.049 	\\ \hline %
sum\_intro		& 0.386			& 0.050 	\\ \hline %
sum			 	& 0.083			& 0.049 	\\ \hline %
tree\_depth		& 0.164			& NI 	\\ \hline %
zip			 	& 1.601			& NI 	\\ \hline %
zip\_unzip		& 1.023			& NI \\%
\hline%
\end{tabular}%
\label{tbl:mochiBM}%
}%
\qquad\qquad%
\subfloat[on \runb{} benchmarks]{%
\begin{tabular}{ |p{0.3\linewidth}|C|C|C| } %
 \hline%
  \emph{benchmark name}		 	&   {\tiny\textsc{Rosette}/ \newline \textsc{Unbound}}	& \mochi \\ \hline%
div-jumps-e				& 0.100			& 1.443 	\\ \hline %
div-jumps				& 0.103			& EOT 	\\ \hline %
fold-append-e			& 0.108			& EOT 	\\ \hline %
fold-append				& 0.120			& EOT 	\\ \hline %
fold-eq-e				& 0.113			& EOT 	\\ \hline %
fold-eq-minus-e			& 0.111			& 0.134 	\\ \hline %
fold-eq-minus			& 0.117			& EOT 	\\ \hline %
fold-eq					& 0.106			& EOT 	\\ \hline %
fold-map-abs-e			& 0.130			& 0.749 	\\ \hline %
fold-map-abs			& 0.121			& EOT 	\\ \hline %
fold-mutations-e		& 0.171			& NI 	\\ \hline %
fold-mutations			& 0.196			& NI 	\\ \hline %
heads-sum-e				& 0.072			& EOT 	\\ \hline %
heads-sum				& 0.076			& EOT 	\\ \hline %
length-append-e			& 0.036			& 0.401 	\\ \hline %
length-append			& 0.019			& EOT 	\\ \hline %
length-append-simpl-e	& 0.035			& 0.978 	\\ \hline %
lucas-vs-fib-e			& 0.134			& 0.073 	\\ \hline %
lucas-vs-fib			& 0.160			& EOT 	\\ \hline %
map-fold				& 0.125			& EOT 	\\ \hline %
mod-div-mult-e			& 0.101			& <1?\footnotemark{}	\\ \hline %
mod-div-mult			& 0.160			& EOT 	\\ \hline %
mutual-recursion-e		& 0.064			& NI 	\\ \hline %
mutual-recursion		& 0.084			& NI 	\\ \hline %
power-monotone-e		& 0.122			& 0.125 	\\ \hline %
power-monotone			& 0.282			& EOT 	\\ \hline %
single-fold-e			& 0.056			& 0.134 	\\ \hline %
single-fold				& 0.063			& 0.239 	\\ \hline %
sorted-e				& 0.163			& NI 	\\ \hline %
sorted					& 0.187			& NI \\%
\hline%
\end{tabular}%
\label{tbl:runbBM}%
}%
}%
\caption{Verification statistics}
\end{figure}%
\footnotetext{The provided error trace was correct, but, due to an internal error, no execution time was reported.}%

Tables~\ref{tbl:runbBM}~and~\ref{tbl:mochiBM} gather statistics on all benchmarks \runb{} was successfully evaluated.
The benchmarks with the names ending with \texttt{-e} are buggy, so an error trace is expected from the verifier.
All other benchmarks are safe.
Two columns of numbers represent execution times (in seconds) in case when the verification process ended successfully by the two competing tools.
\texttt{NI} means that some feature required for verification of the benchmark is not implemented in corresponding tool (see Sect.~\ref{sec:eval}).
\texttt{EOT} denotes that tool terminated with an error or timeout was reached.

Note that the timing were collected on the different machines.
We ran \textsc{Rosette}/ \textsc{Unbound} on Intel(R) Core(TM) i5-6200U CPU @ 2.30GHz, 2 cores, 8 GB RAM.
No desktop version of \mochi{} is publicly available, and thus we we able to ran it only at the web interface (\url{http://www.kb.is.s.u-tokyo.ac.jp/\textasciitilde{}ryosuke/mochi/}), and unfortunately its characteristics are not known to us.

\section{Running \runb{} via DrRacket}

\begin{figure}
  \centering
  \resizebox{0.9\linewidth}{!}{
    \includegraphics{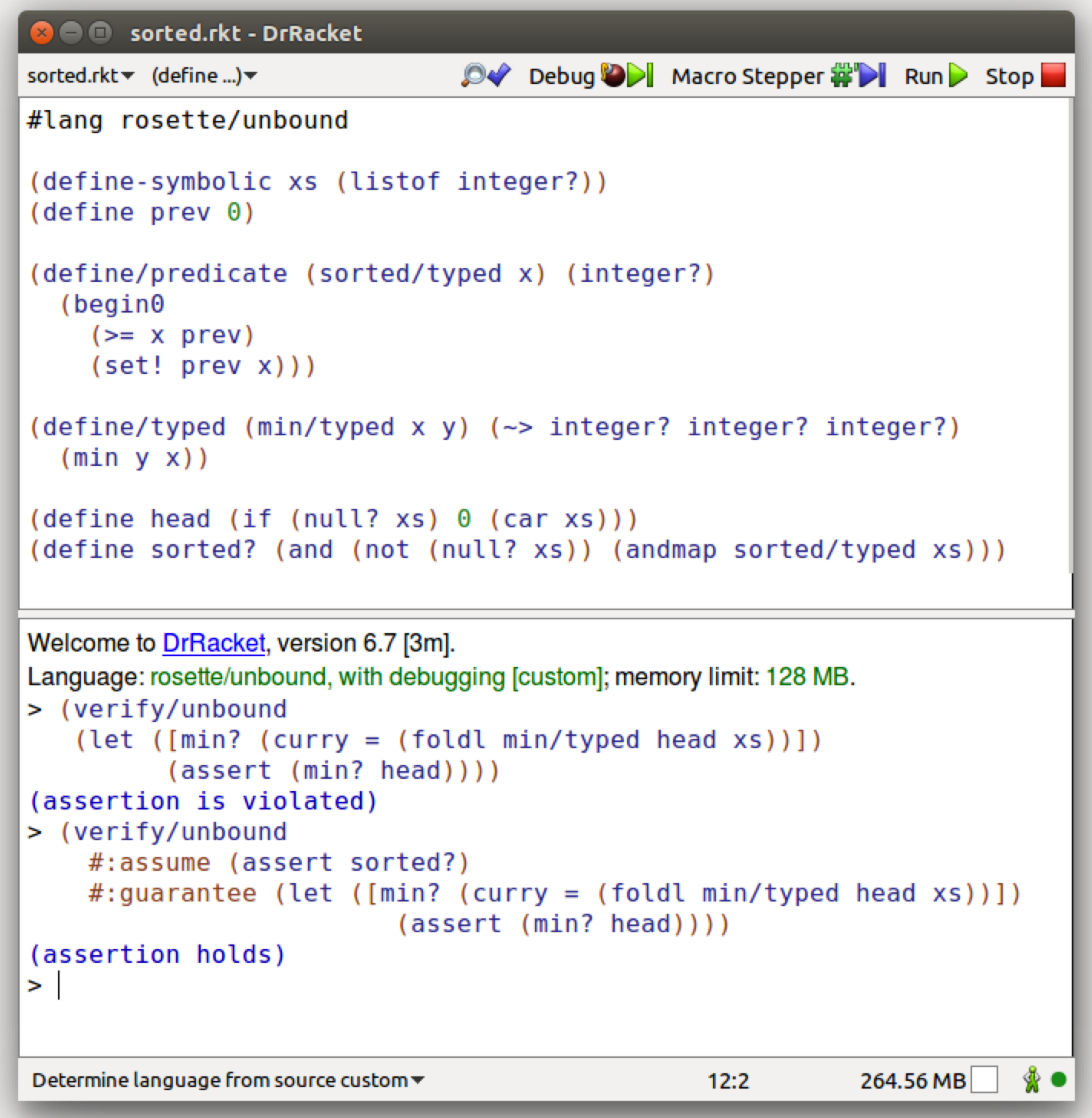}
  }
\end{figure}

\end{document}